\documentclass[%
 reprint,
 amsmath,amssymb,
 aps,
]{revtex4-2}

\usepackage{graphicx}
\usepackage{dcolumn}
\usepackage{bm}
\usepackage{tabularx}
\usepackage{tabulary}
\usepackage{multirow}
\usepackage{siunitx}
\usepackage{comment}
\usepackage{breqn}
\usepackage{inputenc}
\usepackage{appendix}
\usepackage{xcolor}

\begin{document}

\preprint{APS/123-QED}
\title{Cluster Formation and Relaxation in Particle Rafts under Uniform Radial Expansion} 
\author{Kh\'a-\^I T\^o}

\affiliation{%
The Department of Physics and the James Franck and Enrico Fermi Institute\, \\
The University of Chicago\, IL 60637
}%

\date{\today}
\begin{abstract}
Particle rafts floating at an air-liquid interface exhibit a variety of behaviors when interfacial flow is introduced.
Motivated by previous experimental observations, the failure pattern of particle rafts under uniform radial expansion is reported in this paper.
The expansion process is specifically designed to expand the system affinely in the radial direction and to keep the velocity gradient constant throughout. 
A strong resemblance to the results of particle rafts under uniform uniaxial expansion~\cite{D2SM01451C} is found.
The size of the cluster emerging as the rafts are pulled apart scales inversely with the pulling velocity.
This is a result of two competing velocities: the inter-particle separation speed provided by the flow and a size-dependent relaxation speed for clustering.
A model, generalized from a one-dimensional linear (in)stability calculation, is in agreement with the failure morphology found for this radially expanded system.
Nonlinear relaxation and particle rearrangement is observed after the initial clustering occurs. This is a feature unique to a two-dimensional system.
With its easily accessible particle dynamics at the microscopic level, this system provides insights into the morphology controlled by two competing mechanisms in two or higher dimensions and across different scales. 

\end{abstract}

\maketitle
\section{Introduction}
Sub-millimeter particles floating at an air-liquid interface can aggregate into particle rafts.
The inter-particle attraction arises from the coupling between the particles and the fluid interface~\cite{paunov1993lateral,vella2005cheerios,PhysRevE.83.051403}.
The resulting particle rafts can be treated like two-dimensional solids~\cite{PhysRevLett.102.138302}.
For example, compressing 
the raft from two of its opposing ends at small strain can induce an instability in which the particles buckle out of the plane~\cite{PhysRevE.86.031402, PhysRevMaterials.1.042601, doi:10.1021/acs.langmuir.5b01652}. Likewise, pulling the raft from its opposing ends quasi-statically can induce ductile deformation~\cite{D0SM00839G}.
The underlying fluid plays a crucial role.  Not only, does it produce the capillary forces that create the inter-particle attractions, it also provides substantial expansion or compression forces when the fluid surface area is varied sufficiently rapidly. Thus the fluid interface is a valuable tool with which to control particle-raft dynamics when the flow is relevant.

Different regimes of particle-raft behavior have been investigated. A previous study~\cite{D2SM01451C} explored the failure of particle rafts under expansion with a uniformly expanding metric in one dimension.  This produced cracks throughout the material.
By keeping the velocity gradient uniform along the extensional axis, 
the failure was found to be homogeneous with no single crack dominating the resultant morphology. A smooth change in the distance between neighboring cracks was observed as the pulling speed was varied; the failure pattern evolved from an almost unperturbed, densely-packed raft at low shear rates to a uniformly distributed set of rifts separating small clusters of particles at high shear rates. 
In these experiments, the orientations of the particle clusters were found to be nearly isotropic; despite having a specific pulling direction, the clusters were oriented with no strong variation with respect to the extension axis.
This suggests that the direction of the applied shear plays no significant role in determining the failure pattern of particle rafts. 

In other studies, the rafts were pulled apart radially so that the extension was in all directions simultaneously. In one case, this expansion was applied by adding surfactant near its center in order to impose a surface-tension gradient across the raft~\cite{Vella_2004,Bandi_2011}. In another study, advection was generated by pumping water into a funnel so that the raft expanded as the water level rose~\cite{KIM201954}.
In both cases, as the systems evolved under the applied forces, wide cracks appeared preferentially near the center of the rafts.
The morphology in these cases is thus significantly different from what was found in the one-dimensional experiments. 

This raises a puzzle as to what aspects of the expansion  gives rise to the differences observed between the radial and linear pulling geometries.
To address this puzzle, this paper examines the failure morphology of particle rafts pulled apart under radial expansion using a protocol that more closely aligns with that used in the one-dimensional experiments.  In particular, the experiments are designed to produce a nearly uniform velocity gradient everywhere on the air-water interface.  I verify the velocity field using Particle Image Velocimetry (PIV).
The results of these experiments are more similar to the one-dimensional pulling studies suggesting the importance of minimizing velocity gradients in the expansion dynamics.

In order to analyze these results, I generalize two dimensions to an (in)stability calculation used to determine the cluster sizes, \textit{i.e.}, the distance between rifts, in the one-dimensional (1D) situation~\cite{D2SM01451C}.  With no further assumptions on the elastic properties of the aggregate, the calculation showed that the morphology was determined by a cross-over phenomenon due to the competition between the rate of separation (caused by the pulling velocity) and the rate of relaxation (caused by the inter-particle interactions) at different cluster sizes.
In this model, the average cluster length scales inversely with the velocity, in good agreement with the experimental results.

In addition to exploring the failure pattern of particle rafts under a different expansion metric, 
the present study is also inspired by the recognition that the study of structure formation under isotropic competing mechanisms is relevant in many systems across different scales~\cite{C6SM02239A}.
At the largest scale, the competition between gravity and cosmological expansion leads to the cluster formation in the universe~\cite{doi:10.1146/annurev-astro-081811-125502}. 
In particle rafts, the capillary attraction is asymptotically inversely proportional to the distance between particles, which has the same form as two-dimensional gravity.
Therefore, working with an radial-expansion metric provides a two-dimensional table-top experiment for studying this type of phenomenon. 
At much smaller scales, the clustering of colloids~\cite{lu2008gelation, https://doi.org/10.1002/adma.200802763} and the microbial growth on liquid substrate~\cite{PhysRevX.9.021058} both have isotropic competing interaction/flow.
Two-dimensional particle aggregates under radial expansion can offer a macroscopic platform (with access to the microscopic entities involved) to study the morphology controlled by these competing processes.

\begin{figure*}[hbt!]
    \centering
    \includegraphics[width=0.95\textwidth]{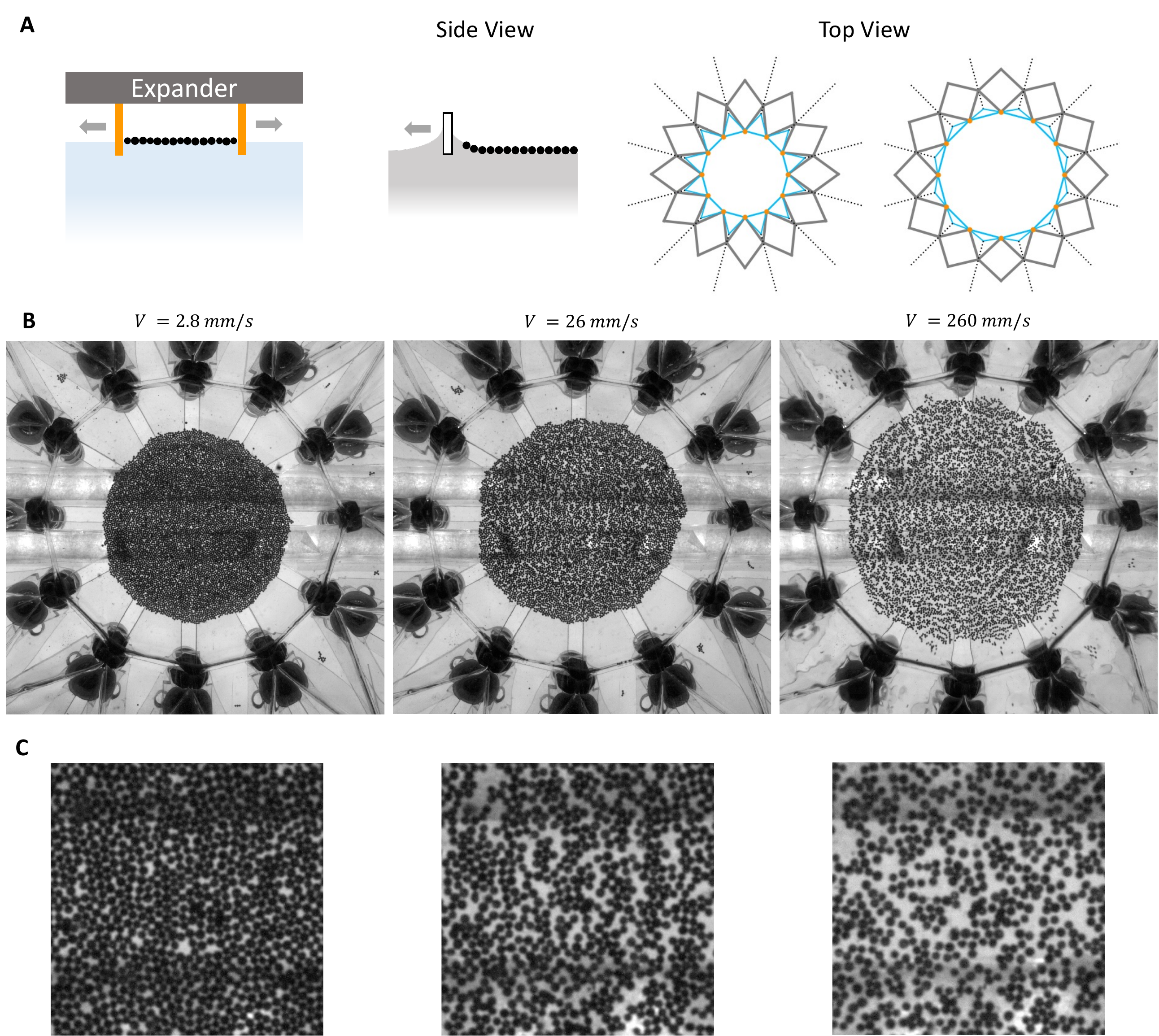}
    \caption{Failure morphology at different pulling velocity $V$. (A) A schematic of the experimental apparatus. Left: The side view of the whole apparatus. A radial expander inserted into water creates an nearly-radial flow to stretch the particle raft enclosed inside. Middle: The side view of the boundary of the Mylar strip boundary. Right: The top view of the expander at early and late time. The gray arms are connected and move coherently to make the twelve orange inner nodes move outward. The inner nodes are connected by Mylar strips painted in blue, which set the boundaries of the system. The dotted black lines show the weight tied to the Mylar strips to keep them tout during the motion. (B, C) Snapshots and zoomed-in images of experiments at a fixed stretch ratio $\lambda = 1.5$ for different pulling velocities $V$.}
    \label{fig1}
\end{figure*}

\section{Experimental Details}\label{sec2}
The particle rafts consist of polyethylene spheres floating at an air-water interface.
The polyethlene particles have density $1080$ $kg m^{-3}$ and are a mixture of two diameter ranges: $d = 550 \pm 50 \mu m$ (small) and $d = 655 \pm 55\mu m$ (large).
Polydisperse packings are made by mixing roughly equal volumes of these two sizes.
The particles naturally aggregate into a raft due to the lateral capillary attraction, also known as the ``Cheerios'' effect~\cite{paunov1993lateral, vella2005cheerios, PhysRevE.83.051403}.
A monolayer of densely-packed particles is prepared by gathering the particles manually into the center of the apparatus as shown in the left panel of  Fig.~\ref{fig1}(A). 
The rafts are compacted by slight vibrations of the interface and have an initial packing fraction of $0.72 \pm 0.01$.  The particle raft floats on deionized water which has density $\rho_w = 998$ $kg  m^{-3}$; dynamic viscosity $\eta_w = 0.95$ $mPa s$ and surface tension, $\sigma_w = 0.073$ $N m^{-1}$ at $22 ^{\circ} C$.

To create an isotropic flow, a two-dimensional radial expander is custom-built to create a coordinated movement of twelve nodes going radially outward simultaneously.  
A schematic of the experimental apparatus is shown in Fig.~\ref{fig1}A.  The nodes are connected by twelve moving Mylar strips to form an expanding twelve-sided polygon.
The Mylar strips are inserted vertically into the water as shown.  Because  Mylar is slightly hydrophilic, 
the particles rafts are enclosed inside the boundaries but do not make \textit{direct} contact with the boundaries.  This is shown in the middle panel of Fig.~\ref{fig1}A.
The rafts are thus pulled only by the flow of the fluid which rises incrementally and uniformly everywhere inside the boundaries as the Mylar boundaries are extended outward radially.  The fluid interface can appropriately be considered as an expanding metric on which the particles float.
As the nodes simultaneously move outward, as shown in ``Top View'' in Fig.~\ref{fig1}A, an approximately isotropic flow is generated by the moving Mylar strips.
(See Appendix A about how Mylar strips are attached and how the expander is driven.)

The distance between the midpoint of each side to the center of the dodecagon is $L_0/2$ and $L_0 \approx 61$ $mm$.
Each side moves outwards away from the center at the same pulling speed $V/2$.
The Particle Image Velocimetry (PIV) measurement at various $V$ is shown in Appendix~\ref{appB}. This allows the flow field  $\boldsymbol{u}(x, y)$  to be described as an affine expansion:
\begin{equation}\label{eq1}
    \boldsymbol{u}(x, y) = \frac{V}{L}(x \boldsymbol{\hat{x}} + y \boldsymbol{\hat{y}})
\end{equation}
where the origin is the center of the raft and $x$ and $y$ are the two orthogonal coordinates in the plane.
The flow is radially outward away from the center.

The pulling speed, $V$, is varied from \SIrange[per-mode=repeated-symbol]{2.8}{260}{\milli\metre\per\second}.
The raft diameter increases as $L = L_0 + Vt$ and the stretch ratio is defined as $\lambda = L/L_0$. The initial rate of stretch, $\dot{\lambda}_0 = V/L_{0}$, is varied from \SIrange[per-mode=repeated-symbol]{5d-2}{4.6}{s^{-1}}. The experiments are stopped at a maximum stretch ratio $\lambda_{max} \sim 1.75$.

I use the method described in~\cite{D2SM01451C} to measure the cluster length, $\ell$, inside each raft.  
This requires counting the number of pixels between rifts on each line of the raster-scanned image. The only complication is due to the difference in pixelization at angles different from the rectangular coordinates of the images produced by the camera.

\section{Experimental Results}\label{sec3}
As the twelve-sided boundary moves outward so that the diameter increases at velocity $V$, the raft of particles is stretched by the underlying flow so that micro-cracks, or rifts, begin to form as shown in Fig.~\ref{fig1}B. 
At intermediate to high $V$, the rifts emerge soon after the onset of expansion.
They are distributed diffusely throughout the entire raft and grow larger with time. 
The experiments are stopped at a stretch ratio ($\lambda_{max} \sim 1.75$).  After their formation and initial growth, the same rifts remain up to a larger stretch ratio; once a rift is formed, it does not collapse at later times. This can be observed in the movies included in the Supplementary Information (SI).
Likewise, the particles in between the initially-formed rifts remain connected.  These clusters of particles maintain their configurations and are distributed homogeneously throughout the entire system with a characteristic size.
The average size of these clusters of particles, $\ell$, is strongly dependent on the pulling speed $V$, as shown in Fig.~\ref{fig1}B, C.

When $V$ is small enough, the raft shape is only very slightly perturbed at the initiation of expansion and remains unchanged afterwards.  There is a small drift in the position due to secondary flows but the internal structure is undisturbed.
In this low-velocity regime, no significant rifts can be observed and the cluster width, $\ell$, remains close to the initial system size, $L_{0}$.  

With increasing $V$, the number of rifts increases while the cluster size, $\ell$, decreases. 
The cluster size, $\ell$ decreases  until it reaches the size of a single particle at large velocities, as shown in the right column of Fig.~\ref{fig1}B.
Figure~\ref{fig1}C shows how the internal features change with increasing $V$: a single relatively closely-packed structure at low $V$, evolves upon increasing $V$ into separate clusters consisting of only a few (sometimes only one) particles.

\begin{figure}[h]
    \centering
    \includegraphics[width=0.48\textwidth]{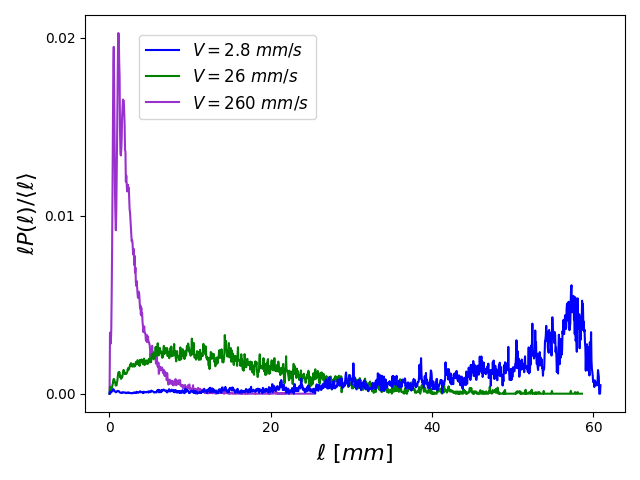}
    \caption{Angle-averaged distribution of cluster lengths, $\ell$, for different velocities at fixed stretch ratio, $\lambda=1.5$.  The most dominant length, that is the probability of cluster length $\ell$ multiplied by $\ell$ normalized by $\langle\ell\rangle = \sum \ell P(\ell)$, is plotted versus $\ell$ with three pulling speeds, $V$: $2.8$ $mm/s$ (blue), $26$ $mm/s$ (green) and $260$ $mm/s$ (purple).}
    \label{fig2}
\end{figure}
The average cluster length, $\langle \ell \rangle$, is obtained by averaging the measurement at all different angles. This is because there is no specific tensile axis in this radial expansion system.
The percentage standard deviation of this average quantity is smaller than $3\%$ for all velocities, which is comparable to the fluctuation due to rotating pixelated images. This indicates that the variation in cluster orientations is very low and below the uncertainty level of detection.

Figure~\ref{fig2} shows the angular-averaged distribution of cluster sizes at a liquid stretch ratio $\lambda=1.5$ for different pulling velocities.
The ordinate is the most dominant cluster length, $\ell P(\ell)/\langle\ell\rangle$, where $P(\ell)$ is the probability of finding a cluster of length $\ell$ and the average cluster length $\langle\ell\rangle = \sum\ell P(\ell)$.
Since the statistics with small cluster lengths always outnumber those with larger $\ell$, I plot the most dominant length, that is how much material has cluster length $\ell$.  This can help better identify the cluster distribution at each velocity, $V$.
At small $V$, a large portion of the distribution remains close to the initial size of the raft, $L_{0}$. 
The distribution shifts to smaller $\ell$ as $V$ increases. 
At the highest pulling speed, $V$, most clusters have lengths between $d$ and $2d$.

\begin{figure}[hbt!]
    \centering
    \includegraphics[width=0.48\textwidth]{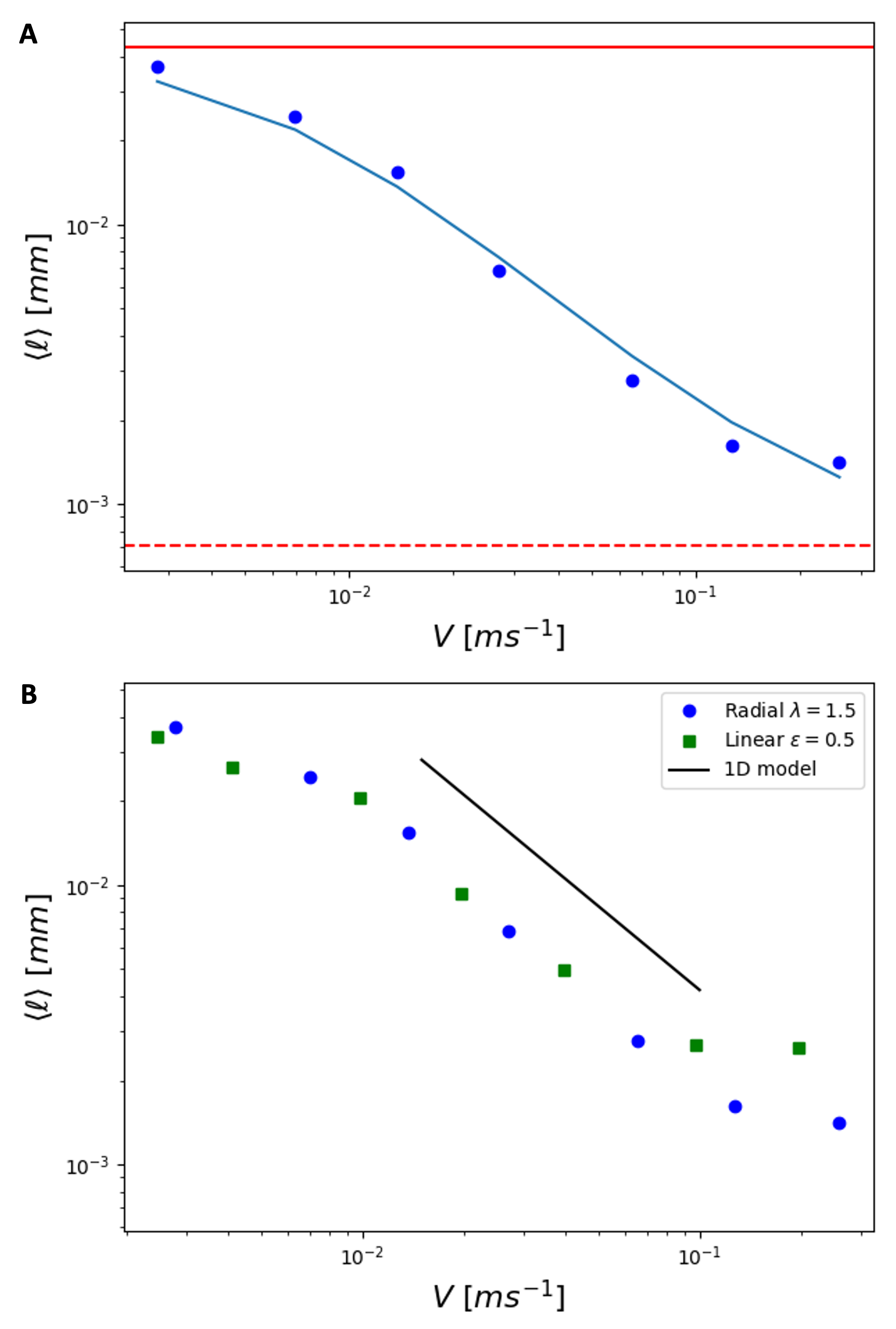}
    \caption{Averaged cluster length, $\langle\ell\rangle = \sum \ell P(\ell)$, versus pulling speed, $V$ at fixed stretch ratio $\lambda=1.5$. (A) The blue points are measurements of $\langle\ell\rangle$ versus $V$. The blue curve is a fit of Eq.~\ref{eq2}. The red solid line indicates $L_{0}/\sqrt{2}$, which is the size of the measuring box. The red dashed line shows the largest particle diameter $d_{max}$.
    The blue curve is a fit to Eq. \ref{eq2}. 
    (B) $\langle\ell\rangle$ is plotted versus $V$ for both radial (blue dots) and uniaxial (green squares) experiments at similar expansion condition ($\lambda = 1.5$ and $\varepsilon = 0.5$).  The black line shows the theoretical prediction for water from Ref. \cite{D2SM01451C}. }
    \label{fig3}
\end{figure}

Figure~\ref{fig3}A shows the average cluster length $\langle\ell\rangle$ versus pulling speed $V$ at a fixed value of $\lambda = 1.5$, for all the experiments.
One can see that $\langle\ell\rangle$ decreases monotonically with increasing $V$ and levels off at high velocity.  
Because a cluster cannot be significantly larger than the initial system size, $L_{0}$, or smaller than a particle diameter, $d$. 
The data can be interpolated between these two extremes using a similar form as was used in the one-dimensional pulling experiments~\cite{D2SM01451C}:
\begin{equation}\label{eq2}
\langle\ell\rangle = \frac{1}{aV^b+1/(L_{0}/\sqrt{2}-d_{max})}+d_{max}. 
\end{equation}
This assumes a power-law dependence of the average cluster length away from the two extremes.
The difference from the one-dimensional fitting form is that initial raft size $L_0$ is scaled by $\sqrt{2}$ because the cluster lengths are measured in a square box of size $L_0/\sqrt{2}$ located in the center of the raft to avoid complexity caused by the round shape of the raft.

The fit gives us $a = (9 \pm 2) \times 10^3$ and $b = 1.20 \pm 0.05$.
This result is similar to the results in uniaxial expansion. The average cluster length, $\langle\ell\rangle$, scales inversely with the pulling velocity $V$.

\section{Discussion}\label{sec4}

Figure~\ref{fig3}B compares the results from these radial expansion experiments with those found in the 1D experiments. Both data sets are shown at a similar time in the expansion: the liquid strain $\varepsilon = 0.5$ for 1D experiment and $\lambda = 1.5$ for radial experiment both occur when the system size is $1.5$ times larger than the original size of the raft.
The experiments under radial, isotropic expansion give very similar results to the 1D experiments. 
This suggests that the experiments with isotropic expansion can in fact be generalized from the analysis used in one-dimension.

To understand this similarity, let us reexamine the flow field described in Eq.~\ref{eq1}.
The velocity field can be expanded around any arbitrary point, $(x_0, y_0)$, within the boundaries:

\begin{equation*}\label{eq3}
    \boldsymbol{u}(x, y) = \boldsymbol{u_{0}}  + \frac{V}{L}((x-x_0) \boldsymbol{\hat{x}} + (y-y_0)\boldsymbol{\hat{y}})
\end{equation*}
where $\boldsymbol{u_{0}}$ is the velocity at $(x_0, y_0)$.
Independent of the position, $(x_0, y_0)$, the velocity field expands radially outward from that point; all points are equivalent and every particle feels the surrounding fluid retreating at the same constant velocity in all directions.  
Thus, the local motion can be reduced to multiple one-dimensional chains as treated in the case of uniaxial expansion.

 The large cracks that were observed near the raft centers in previous experiments where the rafts were pulled radially in two-dimensions~\cite{Vella_2004,Bandi_2011,KIM201954}, are not present in the data presented here.  The difference in the way the pulling of the raft is accomplished is likely the reason for this difference. The present experiment was designed to minimize higher-order gradients in the flow. When using the funnel method~\cite{KIM201954}, I also observed the formation of large cracks which I attributed to gradients in the velocity field produced by the influx of liquid near the center of the raft.  

\begin{figure}[hbt!]
    \centering
    \includegraphics[width=0.45\textwidth]{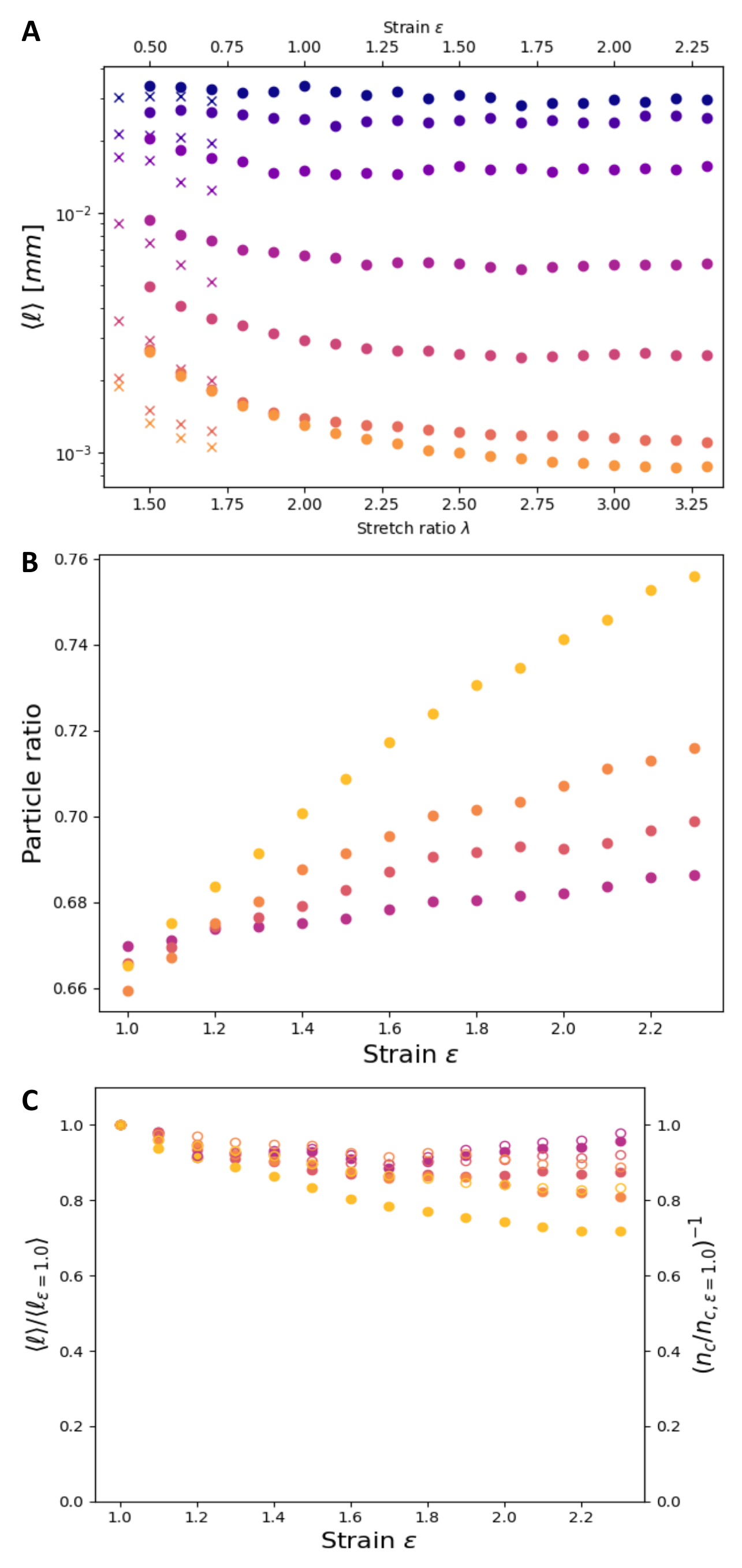}
    \caption{Time evolution of cluster sizes in both radial and uniaxial experiments. (A) Average cluster length, $\langle\ell\rangle$, versus liquid stretch ratio $\lambda$ or strain $\varepsilon$ at all $V$. The results from radial and uniaxial expansion experiments are shown in crosses and dots respectively. The color from dark purple to light orange represent results from fast to slow $V$, and span the entire pulling velocity range in both experiments.) (B) The ratio of particles occupied in the cluster versus strain $\varepsilon$ for uniaxial experiments. The results from four fastest pulling speeds, $V=20$ $mm/s$ (orange) to $200$ $mm/s$ (purple), are presented. (C) The change in $\langle\ell\rangle$ , divided by $\langle\ell\rangle$ at $\varepsilon=1.0$ (solid circles), and the inverse of normalized number of clusters, $n_c$, scaled by $n_c$ at $\varepsilon=1.0$ (empty circles), are plotted versus $\varepsilon$.}
    \label{fig4}
\end{figure}
To understand the two-dimensional effects intrinsic to this systems, I analyze both the uniaxial and radial expansion datasets. 
As shown in Fig.~\ref{fig4}A, the time evolution of the cluster sizes shows a similar trend in both cases.
Here I focus on the time evolution in the uniaxial expansion experiments because they have the greater range. Only the behavior for liquid strain $\varepsilon \ge 1.0$ are examined because the cluster sizes at early times is complicated by limitations of image processing. 
A square observation window of size $L_{x0}$ located at the center of the rafts is also chosen to avoid dealing with the irregular edges of the rafts.

At low $V$, the cluster size is nearly independent of $\varepsilon$.  At higher pulling speed, however, the average cluster sizes consistently decrease with increasing $\varepsilon$.
This suggests non-linear, late-time relaxation of the raft.

One possibility for this is that, after the initial expansion, the particles which are further apart experience a smaller velocity gradient.  However, the particles that are initially more compact continue to experience strong attraction from their neighbors; they rearrange so the particle clusters become more compact.
Another possibility is that the clusters simply continue to break up as time goes on. 

To understand the cause of the decrease in cluster sizes, I calculate the particle ratio inside the clusters that have been identified. 
The particle ratio is defined as the area of \textit{particles} inside the clusters divided by the \textit{total} area of the clusters. This quantity indicates if the voids inside the clusters are shrinking in time. As shown in Fig.~\ref{fig4}B, for four higher pulling velocities, the particle ratios all increase in time. This suggests that rearrangement of neighboring particles inside a cluster is indeed taking place. 

On the other hand, Fig.~\ref{fig4}C examines whether the clusters are breaking up with time. The solid dots represent the cluster lengths $\langle \ell \rangle$ normalized by $\langle \ell_{\varepsilon = 1.0} \rangle$, the measurement at $\varepsilon = 1.0$ versus strain. 
I also plot the inverse of the normalized number of clusters, $n_c$ (scaled by $n_{c, \varepsilon=1.0}$) as empty circles.
$n_c$ is the number of clusters in the observation region normalized by the total area of particles. This quantity shows how many clusters are formed per unit area of materials. If the inverse of $n_c$ matches the change in cluster length, this implies that the change in the number of clusters is the main contributor to the change in cluster sizes, which is the case at $V=20$ $mm/s$
Moreover, the overall change in $\langle\ell\rangle$ is small because the particles are already at a lower-energy configuration to start with. The slight increase in $n_c^{-1}$ after $\varepsilon=1.7$, that is the decrease in the number of clusters, suggests that a larger-scale coalescence has occurred as the global shear became very weak.

At $V=200$ $mm/s$, a significant difference between $n_c^{-1}$ and $\langle \ell \rangle$ can be observed. The average cluster length decreases by more than $20 \%$ while the inverse of $n_c$ only changed by $10 \%$. A substantial rearrangement between neighboring particles is therefore responsible for this change, as shown in  Fig.~\ref{fig4}B as well.

\section{Conclusions}
In this paper, the morphology of particle rafts floating at an air-water interface is studied as the underlying fluid is forced to flow radially outward at a uniform constant velocity gradient.  The experiments are designed to minimize any high-order velocity gradients in the system. By examining the flow field of the radial expansion with particle image velocimetry (PIV), the local expansion metric is found to be uniform inside the apparatus.  
As was found in the case of uniaxial expansion described in Ref.~\cite{D2SM01451C}, the failure morphology due to the outward flow varies smoothly as a function of the fluid expansion velocity.
When the velocity is low, the rafts remain intact throughout the expansion without the formation of visible internal cracks. For larger expansion velocities, rifts open up inside the rafts.   
The rifts are distributed diffusely with a spacing between them that decreases with increasing velocity.  At the largest velocities used in the experiment, the separation decreases until the cluster size, that is the length between rifts, becomes only slightly larger than the size of a single particle.  
The measured average cluster sizes are very close to that of the one-dimensional expansion experiment and follow the same scaling behavior despite the fact that the geometry of the expansion metric is different.
This indicates that the 1D (in)stability model that was used to describe the relaxation and cluster formation in~\cite{D2SM01451C} can be generalized to two-dimensions and is sufficient to explain the radial experiment.

By measuring the time evolution of the clusters, a non-linear relaxation is found at a later time; after the initial formation of clusters, the clusters start to shrink in sizes.
The decrease in particle ratio inside the clusters shows that the particles continue to rearrange and become more compact even while the overall outward expansion continues.
This can be understood because, once the initial expansion rate sets the dominant cluster length, the velocity gradient between clusters is no longer that used initially.  Because the particles are farther apart (in the rifts), the rifts will increase in size faster than the spaces between the more densely packed particles.  Moreover, the global velocity gradient is also decreasing monotonically as the system size becomes larger.
The particles inside an individual cluster will rearrange according to the competition between the local relaxation rate and the local shear rate set by neighboring particles.  This later-time aspect of the relaxation is not captured by the one-dimensional linear (in)stability analysis.

This experiment provides a flexible platform for studying how morphology can form by the competition between two isotropic mechanisms.  There are many other systems that share this common attribute.
As pointed out in~\cite{D2SM01451C}, an obvious analog is cosmological expansion \cite{doi:10.1146/annurev-astro-081811-125502}.  In this table top experiment, the interaction between particles has the same form as gravity in two dimensions.  
Other examples include gel formation\cite{lu2008gelation} and structure formation by colloids in polymer solutions\cite{https://doi.org/10.1002/adma.200802763}.
This two dimensional system gives us a direct access to the clustering morphology with minimal imaging complexity which would be troublesome in three dimensional systems.
In addition, without tuning the details of the potentials chemically or at a small scale, one of the competing processes can be easily manipulated by changing the mechanical driving protocol, while the other can be controlled by changing specific particle-liquid or particle-particle interactions.

\section{Acknowledgement}
I thank T. A. Witten, V. Vitelli and M. M. Bandi for insightful discussion.
I am deeply grateful to S. R. Nagel for his support and mentorship.
This work was supported by the National Science Foundation (MRSEC program NSF-DMR 2011854), by the Simons Foundation for the collaboration Cracking the Glass Problem Award \#348125 and by Government Scholarship to Study Abroad by the Ministry of Education in Taiwan.

\appendix

\section{Details of the Radial Expander}\label{appA}
\begin{figure}[h]
    \centering
    \includegraphics[width=0.48\textwidth]{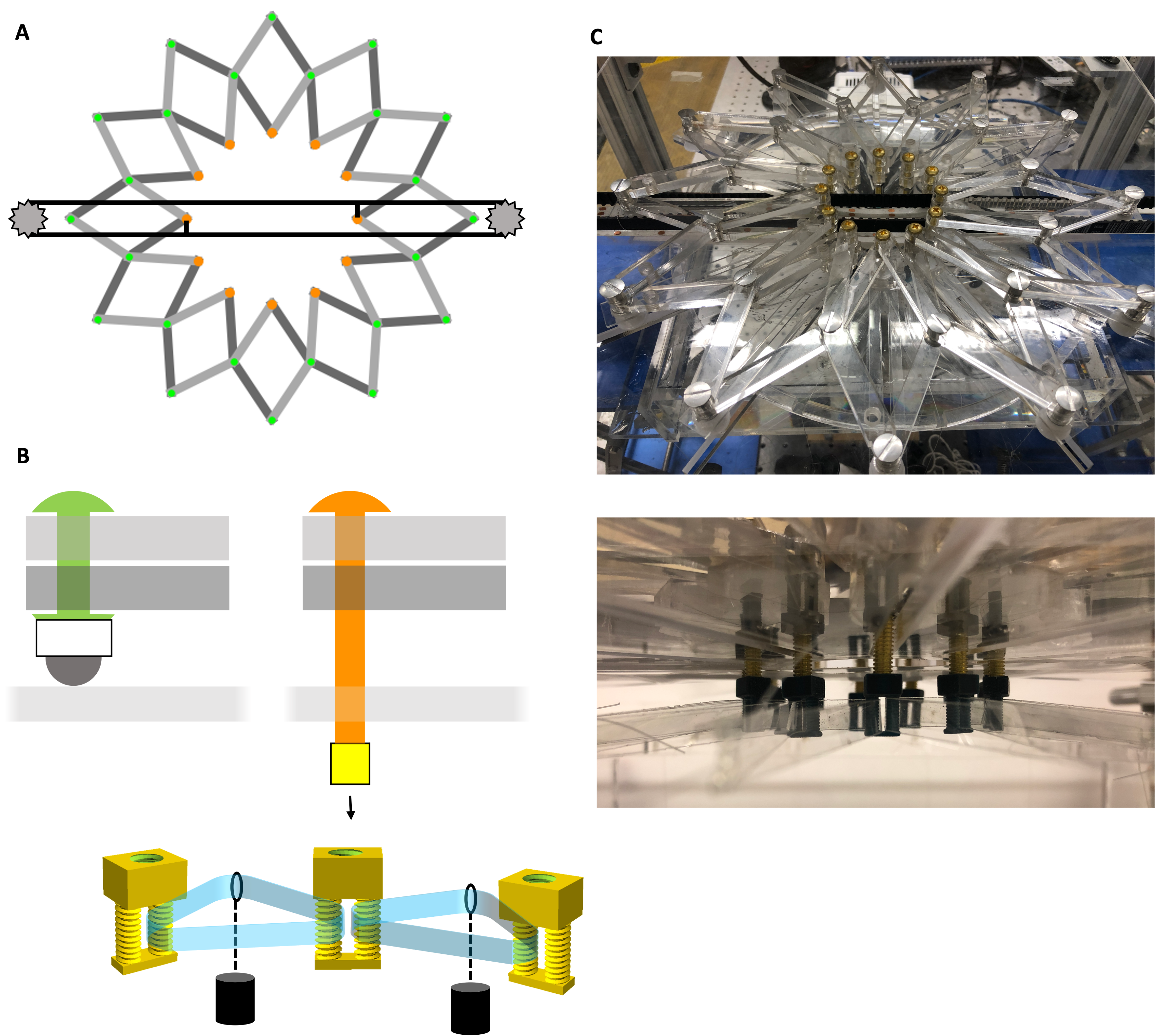}
    \caption{Details of the radial expander design. (A) Schematics of the top view of the radial expander. 
    The inner nodes are shown as orange and the outer nodes are green. Two of the inner (orange) nodes are connected to the opposite sides of a belt which is driven by a motor.  The movement of the belt drives those two nodes in opposite directions.  This opens (closes) the entire structure when the belt is rotated clockwise (counter-clockwise). Because of the geometry connecting all the nodes, pulling on just two of the inner nodes is sufficient for the entire structure to open smoothly.
    (B) A side view of the green and orange nodes. The green nodes are placed to ensure that the entire structure is stable and does not wobble vertically during the opening of the expander.  The bottom of a green node is attached to a ball bearing so that it slides smoothly on a clear acrylic platform. Each orange node is a screw with a 3D-printed attachment at its tip.
    This attachment is composed of two closely-spaced columns that are threaded by two adjacent Mylar ribbons.  These ribbons slide smoothly around the two columns as the expander is moving.  Each ribbon is held in place by the attachments at two adjacent orange nodes.  To keep the Mylar ribbon taut during movement of the expander, a weight is attached that pulls the ribbon outwards on the opposite side from where the particles are located.  The Mylar ribbons are partly submerged below the liquid surface.  It is these ribbons which create the boundary for the particles as shown in Fig.~\ref{fig1}A.
    (C) Top: A photograph of the angled top view of the expander. Bottom: The side view of the chips and Mylar strips before inserting into water.
    }
    \label{figRadialexpander}
\end{figure}
The design of the radial expander used to create the flow with minimal velocity gradients is shown in Fig.~\ref{figRadialexpander}. The twenty-four arms are connected by binding posts (green nodes), which allow them to rotate freely, as shown in the gray pieces in the Fig.~\ref{figRadialexpander}A. Half of them are on the top (light gray) and half of them are at the bottom (dark gray).
Two of the opposite inner nodes are each connected to the opposite side of a belt.
As the belt is driven by a motor, a  motion of the arms to move the twelve inner (orange) nodes outward is created.
Ball bearings are glued underneath the binding posts (shown in Fig. \ref{figRadialexpander}B) to hold the expander and make it glide smoothly on the platform. The inner nodes are long screws, that go below the platform and a 3D-printed chip is attached to the tip. This chip, as shown in Fig. \ref{figRadialexpander}B, has two threaded cylinders with a $1.22$ $mm$ slit in between them.
Each of the cylinder serves as a column for one side of the Mylar strip.
Each Mylar strip forms a triangle and is attached to a weight at the corner against the side connecting adjacent nodes. This provides sufficient tension to the strip during the motion to create an outward-going boundary.

\section{Particle Image Velocimetry}\label{appB}
\begin{figure}[h]
    \centering 
    \includegraphics[width=0.48\textwidth]{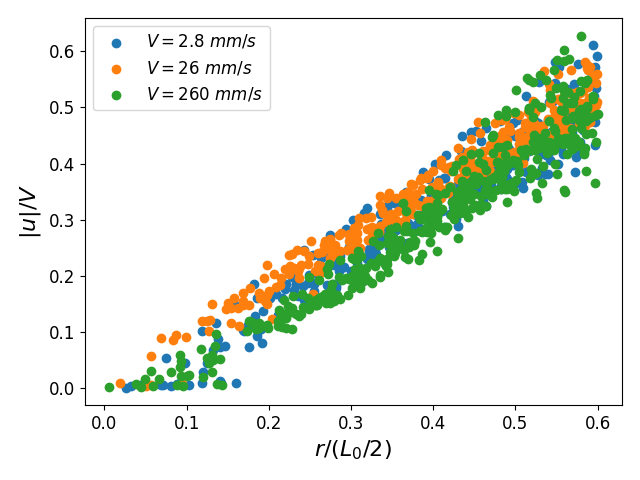}
    \caption{The Particle Image Velocimetry results of the radial expander across whole velocity range. 
    The magnitude of the velocity $\boldsymbol{u}$ normalized by the pulling speed $V$ is plotted versus the position relative to the center, normalized by the radius of the raft, $L_0/2$. A linear relation can be observed in all $V$, indicating an radially affine flow field can be created by the expander.
    The measurements near the boundaries are excluded due to the difficulty of image processing.
    }
    \label{figPIV}
\end{figure}
To measure the expansion of the liquid surface due to the extension of the pullers, the Particle Image Velocimetry (PIV) measurement on water at $V = 2.8$, $26$ and $260$ $mm/s$. The result is shown in Fig. \ref{figPIV}. 
A layer of light, non-interacting floating particles are spread sparsely on the water surface prior to expansion.  By tracking the motion of these particles and computing the correlation between adjacent frames, the underlying fluid flow is shown to  lead to an uniform radial expansion of the surface, as described in \ref{eq1}; the spacing of the particles increases proportional to their relative position.  
Fig.~\ref{figPIV} shows the gradient of the velocity is constant radially.


%

\end{document}